\documentclass[twocolumn]{aastex631}
\usepackage{comment}

\received{October 12, 2021}
\revised{November 8, 2021}
\accepted{November 10, 2021}

\submitjournal{The Astronomical Journal}

\shorttitle{V471 Tau Revised}
\shortauthors{Muirhead et al.}

\graphicspath{{./}{figures/}}

\begin{document}

\title{Revised Stellar Parameters for V471 Tau, A Post-common Envelope Binary in the Hyades}

\author[0000-0002-0638-8822]{Philip S. Muirhead}
\affiliation{Institute for Astrophysical Research and Department of Astronomy, Boston University, 
725 Commonwealth Ave., Boston, MA 02215, USA}
\affiliation{Center for Interdisciplinary Exploration and Research in Astrophysics (CIERA)
and
Department of Physics and Astronomy, Northwestern University, 1800 Sherman Ave., Evanston, IL 60201, USA}

\author{Jason Nordhaus}
\affiliation{Rochester Institute of Technology, National Technical Institute for the Deaf, Rochester, NY 14623, USA}
\affiliation{Rochester Institute of Technology, Center for Computational Relativity and Gravitation, Rochester, NY 14623, USA}

\author[0000-0001-7081-0082]{Maria R. Drout}
\affiliation{University of Toronto, Dunlap Institute for Astronomy \& Astrophysics, Toronto, Ontario M5S 3H4, Canada}

\begin{abstract}

V471 Tau is a post-common-envelope binary consisting of an eclipsing DA white dwarf and a K-type main-sequence star in the Hyades star cluster.  We analyzed publicly available photometry and spectroscopy of V471 Tau to revise the stellar and orbital parameters of the system.  We used archival K2 photometry, archival Hubble Space Telescope spectroscopy, and published radial-velocity measurements of the K-type star.  Employing Gaussian processes to fit for rotational modulation of the system flux by the main-sequence star, we recovered the transits of the white dwarf in front of the main-sequence star for the first time.  The transits are shallower than would be expected from purely geometric occultations owing to gravitational microlensing during transit, which places an additional constraint on the white-dwarf mass.  Our revised mass and radius for the main-sequence star is consistent with single-star evolutionary models given the age and metallicity of the Hyades.  However, as noted previously in the literature, the white dwarf is too massive and too hot to be the result of single-star evolution given the age of the Hyades, and may be the product of a merger scenario.  We independently estimate the conditions of the system at the time of common envelope that would result in the measured orbital parameters today.

\end{abstract}


\section{Introduction} \label{sec:intro}

Post-common-envelope (CE) systems consist of a compact object and a companion in a short-period orbit.  For post-CE systems containing a white dwarf and a main-sequence star, such as V471 Tau, day and subday orbital periods are common \citep{Gomez-Moran2011}.  Initially, two coeval main-sequence stars orbited with a much wider separation.  As the more massive primary star experiences radial expansion during post-main-sequence evolution, the less massive companion can become engulfed in the primary's envelope.  Orbital decay during the common envelope phase significantly reduces the orbital period, until the envelope is ejected leaving a white dwarf and main-sequence companion in a short-period orbit \citep{2006MNRAS.370.2004N,2013A&ARv..21...59I,2019MNRAS.485.4492W}.

V471 Tau is a short-period eclipsing binary containing a hot white dwarf and a K-type main sequence star with kinematics consistent with the Hyades star cluster.  It was first shown to be a spectroscopic binary by \citet{Wilson1953}, and shown to eclipse by \citep{Nelson1970}.  \citet{Young1971} showed that the system is part of the Hyades cluster, and since then the system has been studied in detail.  Once thought to host a circumbinary brown dwarf from eclipse-timing observations \citep{Lohsen1974, Beavers1986, Guinan2001}, high-contrast imaging observations rule out the presence of a massive gravitationally bound third object \citep{Hardy2015}, suggesting that the Applegate mechanism may be responsible for the observed eclipse-timing variations \citep[][]{Applegate1992}.  Observational data of V471 Tau has been collected frequency since its discovery, with multiple teams measuring and fitting the component radial velocities and eclipse light curves \citep{Young1976, Rucinski1981, Bois1988, Ibanoglu2005, Hussain2006, Kaminski2007,  Vaccaro2015} and analyzing far-UV spectra of the white-dwarf component \citep[][]{Barstow1997,Werner1997,OBrien2001,Sion2012}.

Notably, \citet{OBrien2001} acquired multiple ultraviolet spectra of Ly$\alpha$ absorption of the white dwarf using the Goddard High Resolution Spectrograph aboard the Hubble Space Telescope (HST).  They measured multiple radial velocities of the white dwarf, which enabled a more precise measurement of component masses and the white-dwarf temperature.  They found the white dwarf to be unusually hot (34,500 K) for its mass (0.84 $M_\sun$) given the turn-off mass and associate age of the Hyades ($\sim$700 Myr), which is inconsistent with the relatively short timescale for evolution of the white-dwarf progenitor ($M>3$ $M_\sun$) and the expected white-dwarf cooling age ($\sim$10 Myr).  To resolve this paradox, \citet{OBrien2001} proposed a triple star scenario involving a merger of two main-sequence stars to form an Algol-type binary followed by a blue straggler, all with the third main-sequence star orbiting further away.  In their proposed scenario, eventually the blue straggler became an asymptotic-giant branch star, and the wide main-sequence star migrated closer during a common envelope phase, eventually resulting in the system as we see it today.  A curious feature of the literature measurements of V471 Tau system is the relatively large radius for the main-sequence star given its mass and the age of the Hyades \citep[$R=0.96 \, R_\sun$, $M=0.93 \, M_\sun$ from][]{OBrien2001}.

Recently, V471 Tau was observed by NASA's K2 Mission in short-cadence mode, providing a high signal-to-noise light curve of the system.  In this paper we present a reanalysis of publicly available data of V471 Tau, including the K2 light curve, to determine the most accurate and precise system parameters. We employed Gaussian processes and affine-invariant Markov chain Monte Carlo methods to simultaneously fit orbital parameters and the main-sequence-star variability \citep{Foreman-Mackey2013, Foreman-Mackey2017}.  Using this approach, we recovered the transit of the white dwarf in front of the main sequence star in the K2 data, which is shallower than would be expected from a purely geometric transit due to the effects of gravitational microlensing.  We incorporated published radial-velocity measurements of each star, and we remeasure radial velocities of the white dwarf component using archival HST spectra.  Unlike previous analyses, we find the mass and radius for the K-type main-sequence star to be fully consistent stellar evolutionary models for the age and metallicity of the Hyades.  Our resulting best-fitting mass for the white dwarf is consistent with previous investigations, and it remains too hot and too massive to be formed by single-star evolution.

This paper is organized as follows: In section \ref{sec:data} we describe the datasets used in this analysis.  In section \ref{sec:fit} we describe our eclipsing binary model and fitted best parameters to the system.  In section \ref{sec:discussion} we discuss implications for the revised parameters on the evolution of the system, and in section \ref{sec:conclusions} we summarize our results.

\section{Data} \label{sec:data}

\subsection{K2 Light Curve}

V471 Tau was observed for roughly 80 days by the K2 Spacecraft during Campaign 4 for Guest Observer programs 4027\footnote{\url{https://keplergo.github.io/KeplerScienceWebsite/data/k2-programs/GO4027.txt}} and 4043.\footnote{\url{https://keplergo.github.io/KeplerScienceWebsite/data/k2-programs/GO4043.txt}}  The observations were acquired in short-cadence mode, with integration times of 58.89 s. Data from NASA's K2 Mission contain instrumental systematic effects due to pointing drift in the space telescope \citep{Howell2014}. To remove the instrumental systematic effects, we used the EPIC Variability Extraction and Removal for Exoplanet Science Targets (\texttt{everest}) software package \citep{Luger2016, Luger2018}.  In addition to the expected instrumental systematic effects, the raw K2 light curve also contains anomalously high- and low-flux measurements occurring at regular intervals.  We identified the period and ephemeris of the anomalous measurements using a Lomb--Scargle periodogram ($P$=1.96066 days).  We used the masking feature in the \texttt{everest} pipeline to exclude the anomalous data points and white-dwarf occultation events from the \texttt{everest} algorithm, as sharp features can adversely affect the quality of the systematic corrections.  Manual inspection of the resulting light curve shows the periodic white-dwarf occultation events, flaring, and quasiperiodic variations  consistent, which are consistent expectations from synchronous rotation and evolving magnetic spots on the main-sequence star.

\subsection{Radial Velocities}

For the radial-velocity measurements of the main-sequence star, we used 202 archival ground-based radial-velocity measurements and uncertainties from the literature \citep{Bois1988}.  The measurements span the entire orbit of the main-sequence star and were reported in heliocentric Julian date.  Radial-velocity measurements of the white dwarf are sparser owing to the high contrast between the white dwarf and the main-sequence star at visible wavelengths.  We used eight published measurements from \citet{OBrien2001} taken with HST-GHRS after the COSTAR upgrade to the telescope.  The measurements used the Stark-broadened wings of the white dwarf's Ly$\alpha$ absorption line to determine the white dwarf's radial velocity.  We note that the zero point for the published white-dwarf radial-velocity measurements is uncertain, and we chose to leave that as a free parameter in our fit to the data.

V471 Tau was also observed by the Space Telescope Imaging Spectrograph (STIS) on the HST, with publicly available reduced and calibrated data \citep{Ayres2010}. Fourteen of the observations were taken using the E140M grating, which includes the white-dwarf Ly$\alpha$ absorption.  The observations were originally collected to measure emission and absorption features in the system by \citet{Sion2012}.  They found that different species of absorption lines show different radial-velocity semi-amplitudes, indicating nonisotropic emission.  We therefore follow the approach of \citet{OBrien2001} and use the Stark-broadened wings of the Ly$\alpha$ absorption line to represent the radial velocity of the white dwarf, excluding regions with other absorption or emission lines.  We note that the STIS spectra are Doppler corrected for spacecraft motions onboard. 

\begin{figure}
    \centering
    \includegraphics[]{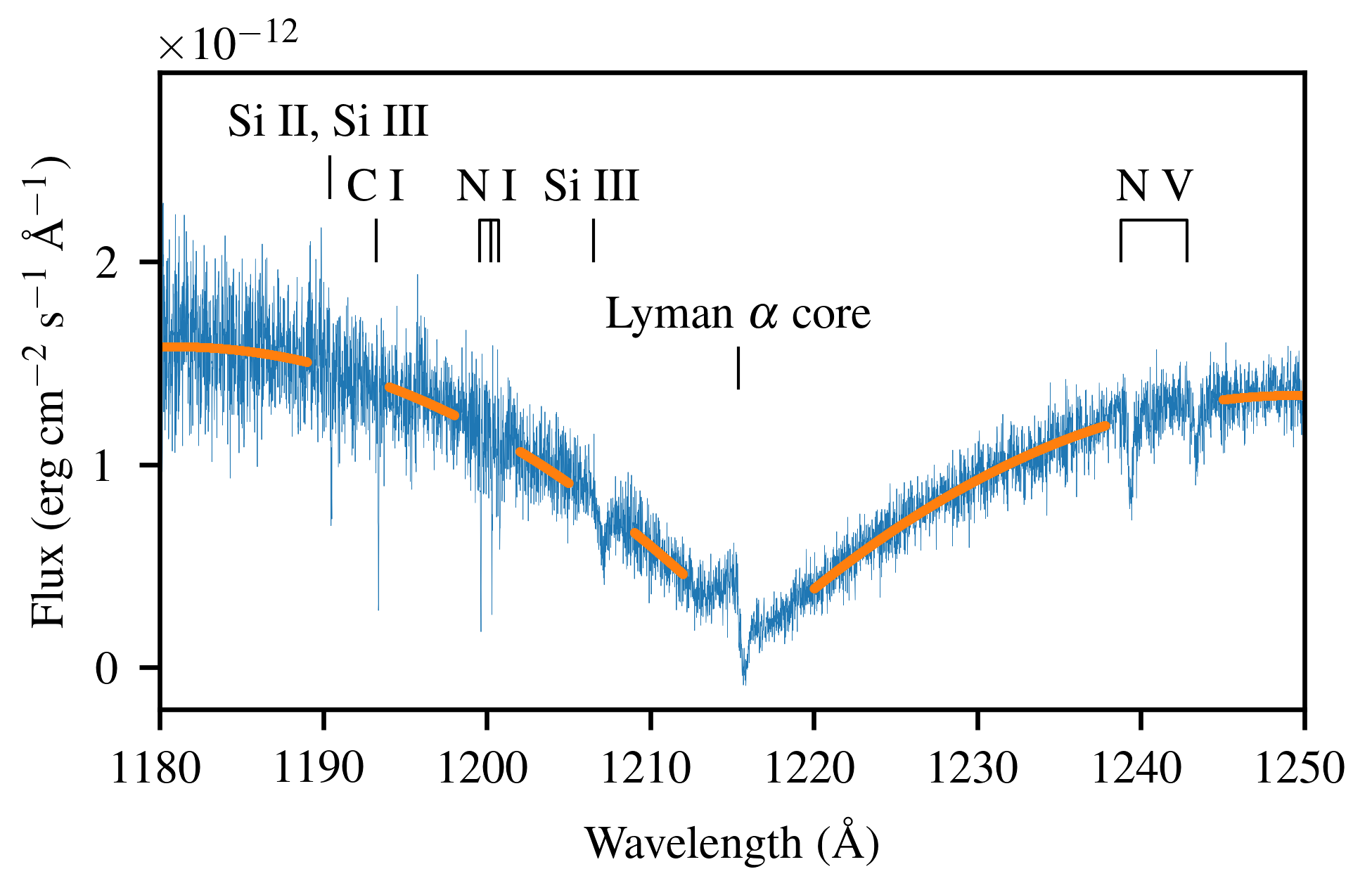}
    \caption{Far-ultraviolet spectrum of V471 Tau.  An archival HST spectra acquired with the Space Telescope Imaging Spectrograph (STIS) is shown in blue.  In orange we show our best-fitting model to the Stark-broadened wings of the white dwarf's Ly$\alpha$ absorption line, ignoring regions contaminated by chromospheric emission and interstellar absorption.}
    \label{fig:stis}
\end{figure}

We originally fit model white-dwarf spectra to the data to determine radial velocities; however, the resulting radial velocities contained an obvious multiplicative bias depending on the choice of model temperature, owing to the temperature-dependent slope of the continuum.  Instead, we constructed a spectral template from the data itself.  We downloaded and coadded the spectra, then fit a piece-wise second-degree polynomial template to each side of the Ly$\alpha$ absorption line, ignoring regions with chromospheric emission, interstellar absorption and geocoronal emission (see Figure \ref{fig:stis}).  We then fit the piece-wise polynomial template to each individual spectrum, allowing only radial velocity of the template and overall normalization to vary with each fit.  Since the template was constructed from the data itself, it cannot serve as an absolute radial-velocity calibration, meaning we cannot measure the systemic radial velocity or the gravitational redshift of the white dwarf component, both of which affect the radial-velocity zero-point.  But, the template can serve to measure relative changes in radial velocity.  The resulting radial velocities clearly follow the expected relative radial-velocity variation due to the white-dwarf orbit. 

Following our approach with the GRS radial-velocity measurements, we treat the zero-point radial velocity of the STIS measurements as another free parameter in the full eclipsing binary fit.  To estimate the uncertainty in the STIS radial-velocity measurements, we fit a simple sinusoid to the measurements versus orbital phase, fixing the sinusoidal period to that of the system and allowing the amplitude and offset to vary.  We calculated the standard deviation between the measurements and the fitted sinusoid as 11.2 km s$^{-1}$.  We adopted that standard deviation as the individual measurement uncertainties in each redial velocity observation.  Table \ref{tab:stis} details the 14 observations and resulting radial-velocity measurements.

We note that since the template was constructed by coadding spectra, there will be some broadening or smoothing of the template due to the radial-velocity differences in those spectra. To test this effect, we recalculated the template, this time shifting the individual spectra to remove the orbital radial-velocity differences before coadding them.  This approach had a negligible impact on the resulting radial velocities and estimated uncertainties, each changing by less than one tenth of the estimated measurement uncertainty.

\begin{deluxetable}{lcc}
\tablecaption{Archival HST-STIS Observations and Radial-velocity Measurements (fitted offset subtracted, $\sigma$=11 km s$^{-1}$)\label{tab:stis}}
\tablewidth{0pt}
\tablehead{
\colhead{Obs. ID} & \colhead{Mid-exposure BJD} & \colhead{RV (km s$^{-1}$)}}
\startdata
o4mu02010 & 2450885.68119 & 59 \\
o4mua2010 & 2450885.74033 & -42 \\
o4mua2020 & 2450885.80550 & -145 \\
o4mu01010 & 2450896.50047 & 154 \\
o4mu01020 & 2450896.55827 & 144 \\
o4mua1010 & 2450896.62779 & 58 \\
o5dma1010 & 2451781.22802 & -140 \\
o5dma2010 & 2451782.29258 & -146 \\
o5dma3010 & 2451784.11080 & 138 \\
o5dma4010 & 2451781.49621 & 160 \\
o6jc01010 & 2452299.57640 & 169 \\
o6jc01020 & 2452299.63568 & 105 \\
o6jc01030 & 2452299.70243 & -43 \\
o6jc01040 & 2452299.76917 & -131 \\
\enddata
\end{deluxetable}

\begin{figure*}
    \centering
    \includegraphics[width=0.9\textwidth]{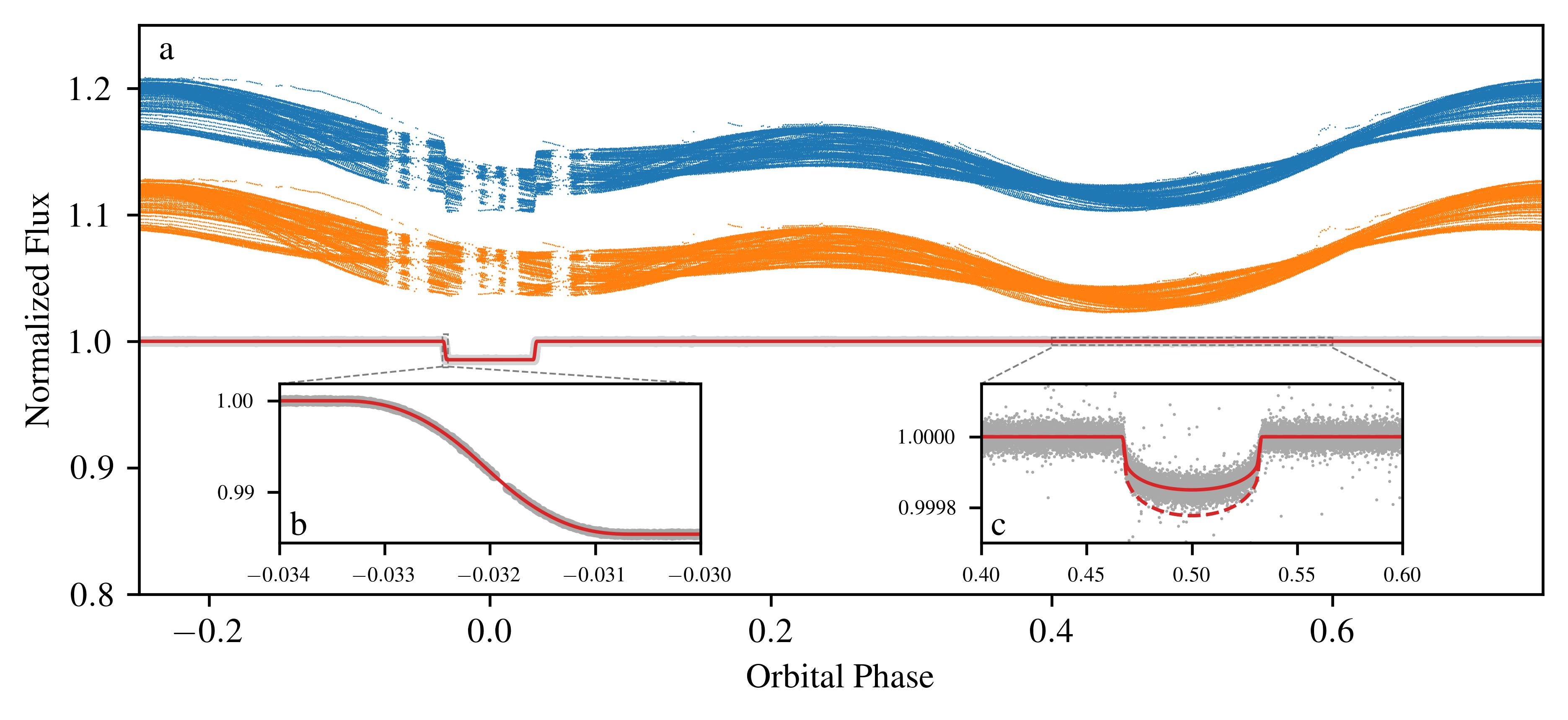}
    \caption{Phase-folded light curve of V471 Tau.  (a) Full orbital phase showing the normalized K2 data (blue, offset for clarity), the best-fitting stellar variability model (orange, offset for clarity), the K2 data divided by the variability model (gray), and the best-fitting eclipse model (red).  Gaps in the data near zero phase correspond to periodic anomalous data points removed from the light curve (see text). (b) Insert showing the ingress of the white-dwarf occultation.  (c) Insert showing the white-dwarf transit and the best-fitting eclipse model with and without gravitational microlensing (red and dashed red, respectively).}
    \label{fig:k2}
\end{figure*}

\begin{figure}
    \centering
    \includegraphics[width=0.47\textwidth]{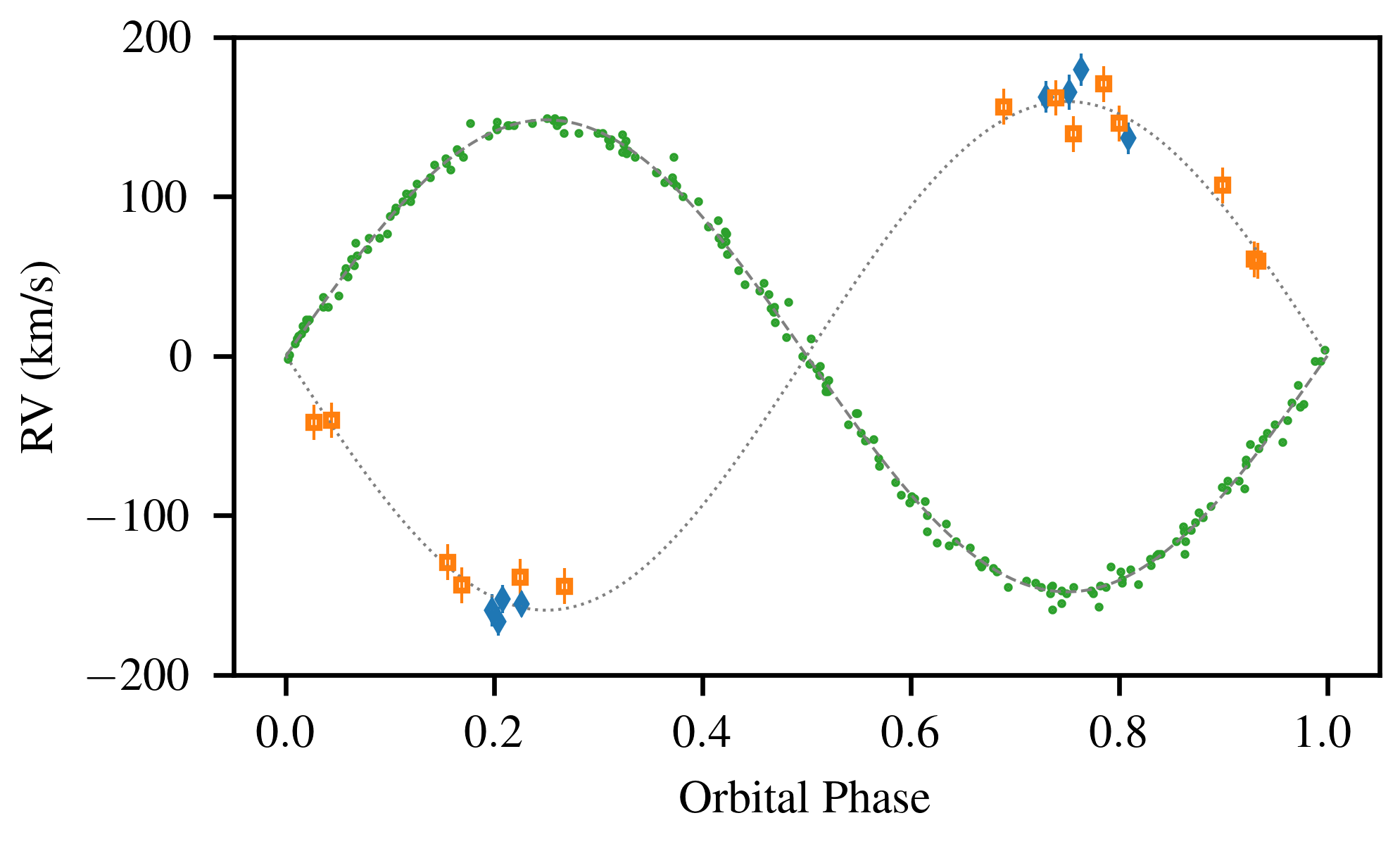}
    \caption{Radial-velocity measurements of the component stars of V471 Tau.  Green circles show measurements of the main sequence star from the literature \citep{Bois1988}, blue diamonds show measurements of the white dwarf from the literature \citep{OBrien2001}, and orange squares show measurements from this effort using a custom reduction of archival HST-STIS observations.  The best-fitting models for the main-sequence star ( solid gray) and white dwarf (dashed gray) are shown.  The fitted radial-velocity zero-points for each data set have been subtracted from the measurements and model.}
    \label{fig:rv}
\end{figure}

\begin{figure*}
    \centering
    \includegraphics[width=\textwidth]{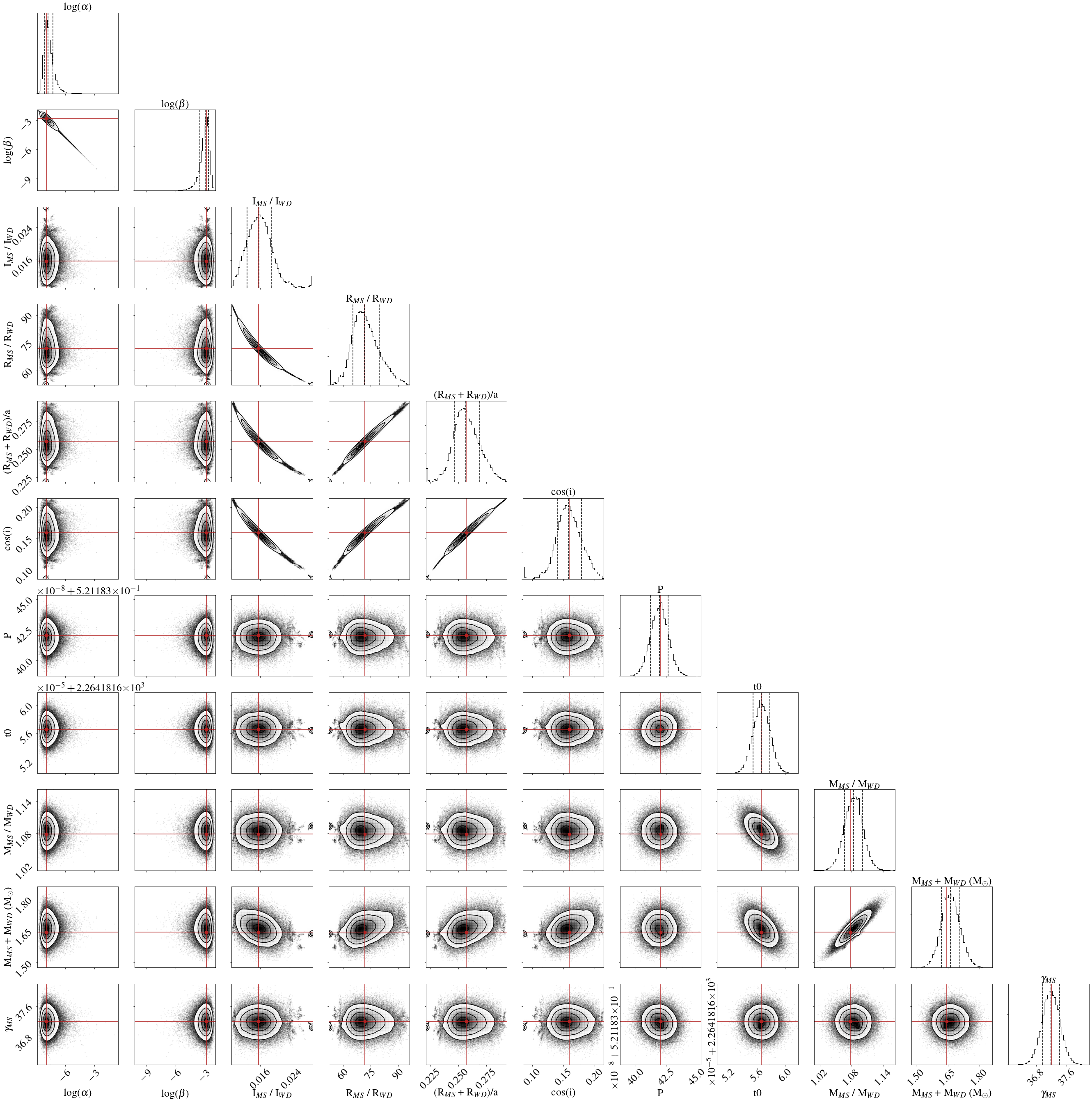}
    \caption{Density of samples of the posterior probability distribution for fitted parameters.  Highest density regions correspond to higher likelihood parameters.  Histograms at the top capture the marginalized posterior probability distribution for a given parameter.  Dashed lines indicate the 16\%, 50\%, and 84\% quantile locations.  Red lines and square indicate the values for the sample with the highest posterior probability. This figure was made using \texttt{corner} \citep{corner}.}
    \label{fig:corner}
\end{figure*}

\begin{figure}
    \centering
    \includegraphics[width=0.47\textwidth]{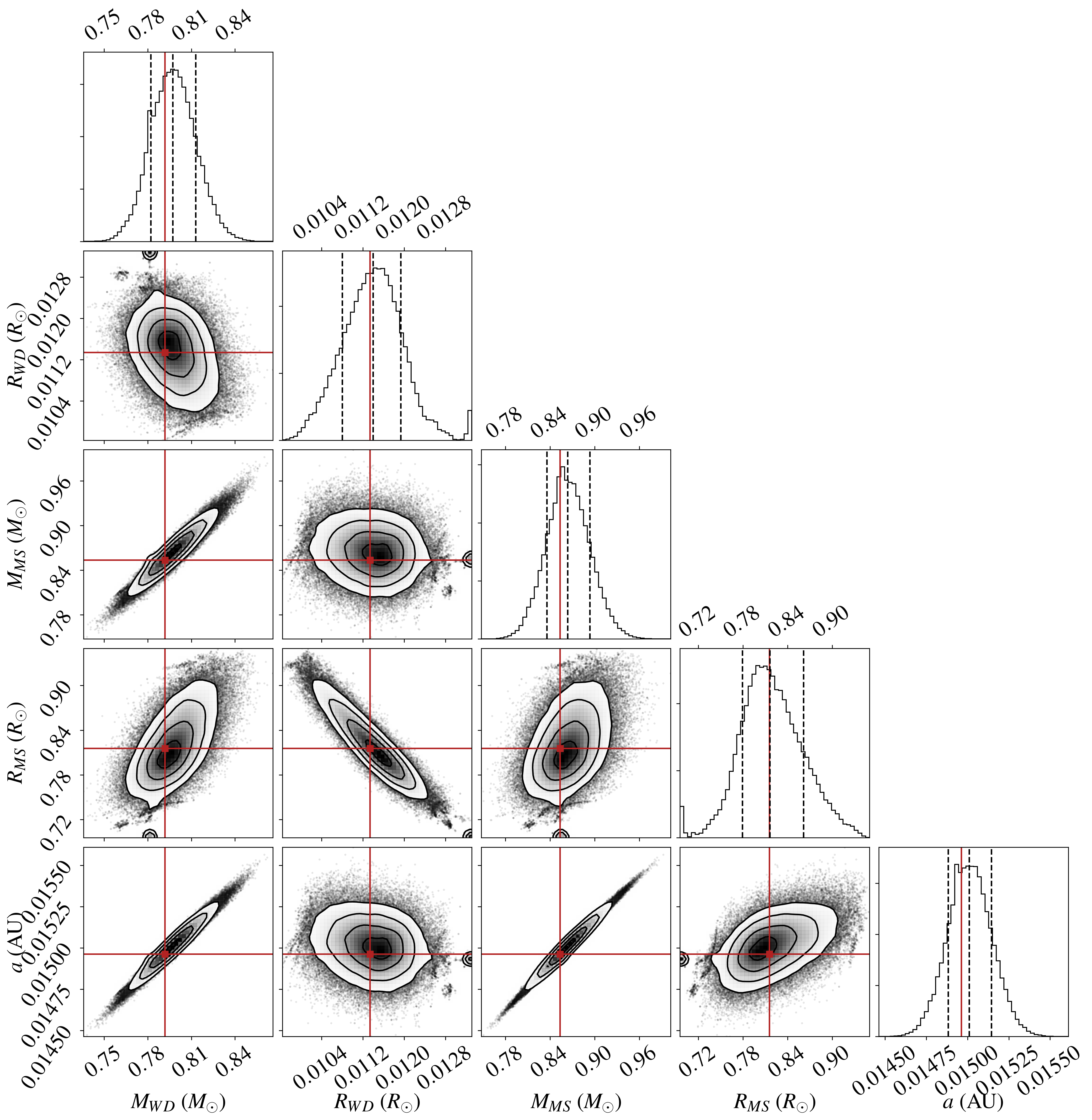}
    \caption{Same as Figure \ref{fig:corner} but for parameters of interest that were derived from the fitted parameters.}
    \label{fig:corner2}
\end{figure}

\section{Fitted Model} \label{sec:fit}

To fit for the system orbital parameters, we combined three publicly available software packages in a Bayesian framework: an eclipsing binary light-curve modeling package called \texttt{eb} \citep{Irwin2011}, a Gaussian processes software package commonly used for fitting rotational variability called \texttt{celerite} \citep{Foreman-Mackey2017}, and a package to sample the resulting posterior probability distribution using affine-invariant Markov chain Monte Carlo methods called \texttt{emcee} \citep{Foreman-Mackey2013}.  In our model, we assumed an orbital eccentricity of zero, given the relatively short timescale for orbital circularization.

\subsection{Eclipsing Binary Model}

The \texttt{eb} software package produces a model light curve for a given set of orbital parameters, and includes light travel time and limb-darkening effects.  We used \texttt{eb} to model the eclipse events, with some custom modifications.  The \texttt{eb} package does not include gravitational microlensing effects during eclipses, which can be significant for the transit of compact objects.  During the eclipse of the K dwarf by the white dwarf (hereafter, the transit), gravitational microlensing will ``fill in'' the transit depth by the differential amount $2R_{\rm Ein}^2 / R_{MS}^2$ \citep{Agol2003}, where $R_{\rm Ein}$ is the Einstein radius of the white dwarf during transit and $R_{MS}$ is the radius of K dwarf. To modify the \texttt{eb} model to incorporate gravitational microlensing, we isolated the region of the model that contained the transit event, subtracted the flux contributed by the white dwarf, renormalized the light curve to a maximum of unity, then scaled the model transit to account for gravitational microlensing. If the Einstein radius is sufficiently large, the filling factor results in an inverted transit  \citep[e.g.][]{Kruse2014}; in the case of V471 Tau,  gravitational microlensing results in a shallower transit, similar to what was seen in the post-common-envelope binary KOI-256 \citep{Muirhead2013}.  For a set of orbital parameters, we computed the model light curve across the full orbital phase with a sampling of roughly 1 s.  We convolved the model light curve with a normalized top hat function of width equal to integration time of the K2 short-cadence data (58.89 s), in order to capture the effect of finite integration time on the light curve.  We then interpolated the model light curve onto the K2 time stamps using cubic spline interpolation.

\subsection{Accounting for Stellar Variability}

The \texttt{celerite} software enables the treatment of stellar variability as correlated noise added to a ``mean model,'' which in this case is the light-curve model produced by \texttt{eb} for a given set of orbital parameters.  In this approach, the stellar variability is treated as correlated noise added to the purely geometric eclipsing binary model.  However, since the eclipse events result in fractional changes in the overall system flux, a correct treatment would involve multiplying the stellar variability model by the eclipse model, rather than adding them.  We therefore chose to fit the log of the eclipse model, with added correlated noise, to the log of the K2 data, propagating the corresponding measurement errors appropriately.  This results in the correct treatment of fractional changes in the system flux.

In a Gaussian-processes approach, correlated noise is described by a user-selected kernel function containing free parameters.  The kernel function relates the degree of correlation between data points, which are then entered into the data covariance matrix for determining the likelihood probability density for a given choice of free parameters (both kernel parameters and mean-model parameters).  Following the approaches described in \citet{Foreman-Mackey2017}, for the kernel, we chose a damped harmonic oscillator with two free parameters: an amplitude $\alpha$ and an damping constant $\beta$.  We fixed the periodic component of the damped harmonic oscillator to one half of the orbital period of the system $P$, which is the typical periodicity of the rotational variations as evidenced by visual inspection:

\begin{equation}
    \kappa(\tau) = \alpha \, e^{-\beta \tau} (1 + \cos{\frac{\pi \tau}{P}})
\end{equation}

\noindent where $\kappa$ is the degree of correlation between K2 data points separated by $\tau$ in time.  

\subsection{Radial-Velocity Model}

To model the radial-velocity observations, we used a sinusoidal model for each component of V471 Tau corresponding to the orbit expected from the fitted orbital parameters.  As noted in the literature, the radial-velocity zero point for the literature white-dwarf measurements is uncertain \citep{Vaccaro2015}.  For the two sets of HST radial velocities of the white dwarf component---the literature measurements from \citet{OBrien2001} using GHRS and our measurements from archival STIS observations---we included a separate additive offset for each to account for the systematic offsets in the radial-velocity zero-points.  

For a given set of orbital parameters, we computed the corresponding sinusoidal radial-velocity variation for each component with a sampling of 1 s.  We convolved the resulting model radial-velocity curve with a tophat of width 1800 s, to account for the typical integration times of the radial-velocity observations and resulting smoothing of the purely geometric radial velocities.  In reality, the integration times for the radial-velocity observations varied; however, the choice of modeled exposure time had a negligible affect on the resulting best fit parameters, so we assume an 1800 s exposure for all radial-velocity observations.

For a given set of model orbital parameters and measured radial velocities and uncertainties, the log of the likelihood probability density was computed and added to the log of the likelihood probability density for the light-curve model. 

\subsection{Fitting Procedure}

To determine the best-fitting kernel parameters and orbital parameters, we sampled a Bayesian posterior probability function assuming Gaussian errors for all the measurements and assuming uniform prior functions on all free parameters with reasonable upper and lower bounds.  For the independent measurement uncertainties on each data point, were used the uncertainties returned from the \texttt{everest} pipeline, the reported uncertainties for the radial-velocity measurements in the literature, and our estimated uncertainties from the STIS measurements.  We assumed an eccentricity of zero, fitting each component's radial-velocity measurements with a sinusoidal function offset by $\pi$ respectively, and we assumed no third light contamination and no reflected light in the eclipse model.   We assumed quadratic limb darkening in both stars, using predicted values for main-sequence dwarfs and white dwarfs with the expected surface gravity and surface temperature in the K2 bandpass \citep{Claret2018,Claret2020}.  We did not include ellipsoidal variations in the mean model, letting the Gaussian processes approach capture all stellar variability effects.  The fitted Roche-lobe filling factor for the main-sequence star is 52\% (here defined as the radius of the star divided by the distance to the L1 Lagrange point), implying that nonspherical effects should have a negligible impact on the mean model.  All in all, we fit for the following parameters: kernel amplitude, kernel damping constant, orbital period, occultation ephemeris, sum of the radii in units of semi-major axis, radius ratio, surface brightness ratio in the K2 band, orbital inclination, sum of the masses, mass ratio, systemic radial velocity of the system, and the radial-velocity zero points for the two sets of white-dwarf RV measurements. 

To find the mostly likely orbital parameters and their uncertainties, we sampled the Bayesian posterior using the \texttt{emcee} software.  We used 100 chains, each starting with random parameters spread across the ranges of their respective uniform prior functions.  We ran the sampling algorithm for 100,000 steps (a ``burn-in'') to find the highest likelihood parameters.  We then divided the data by the best-fitting eclipse model and smoothed the resulting light curve with a median filter of with 61 data points (corresponding to roughly 1-hour of observations).  Data points that differed from the smoothed light curve by more than three sigma were removed from the light curve, and ran another 100,000-step burn-in.  This step removed the flare events in the light curve.  Following the second burn-in, we removed chains with maximum posterior values that were significantly lower than the highest posterior value across all chains (5 our of 100 chains), and ran a 200,000-step ``production'' to sample the peak of the posterior probability density function as a function of free parameters.  

Figure \ref{fig:k2} shows the phase-folded K2 light curve and best-fitting model: that is, the sample of free parameters with the highest posterior probability.  We recover the transit of the white dwarf in front of the main-sequence star, with a pronounced effect from gravitational microlensing.  Figure \ref{fig:rv} shows the radial-velocity measurements and the best-fitting model.  Figures \ref{fig:corner} and \ref{fig:corner2} show the sampling distributions for the fitted and derived parameters, respectively.  Table \ref{tab:params} lists the fitted parameters corresponding to the sample with the highest posterior probability density, the median of the MCMC steps, and the standard deviations of the parameter samples, which serve as an estimate for the marginalized uncertainty in each parameter.  Table \ref{tab:params} also lists several several system parameters derived from the fitted parameters, namely the individual stellar masses and radii.

\begin{deluxetable*}{rcccl}
\tablecaption{Revised parameters for V471 Tau\label{tab:params}}
\tablewidth{0pt}
\tablehead{
\colhead{Parameter} & \colhead{Max. Posterior} & \colhead{Median of MCMC} & \colhead{Std. Dev. of MCMC} & \colhead{Notes/Units} \\
\hline
\multicolumn5c{Fitted parameters}
}
\startdata
log($\alpha$) & -7.95 & -7.81 & $\pm$ 0.53 & Kernel parameter\\
log($\beta$) & -2.83 & -2.97 & $\pm$ 0.53 & Kernel parameter\\
$I_{MS}$ / $I_{WD}$ & 0.0155 & 0.0157 & $\pm$ 0.0033 & Brightness ratio in the K2 band\\
$R_{MS}$ / $R_{WD}$ & 71.9 & 71.5 & $\pm$ 7.3 & Radius ratio\\
($R_{MS}$ + $R_{WD}$)/$a$ & 0.257 & 0.257 & $\pm$ 0.012 & Sum of radii over semi-major axis\\
cos($i$) & 0.159 & 0.158 & $\pm$ 0.020 & Cosine of the inclination\\
$P$ & 0.5211834204 & 0.5211834194 & $\pm$ 0.0000000072 & Orbital period (days)\\
$t_0$ & 2264.1816566 & 2264.1816566 & $\pm$ 0.0000012 & Occultation ephemeris (BJD - 2454833.0)\\
$M_{MS}$ / $M_{WD}$ & 1.078 & 1.084 & $\pm$ 0.017 & Mass ratio\\
$M_{MS}$ + $M_{WD}$ ($M_{\odot}$) & 1.645 & 1.661 & $\pm$ 0.044 & Mass sum ($M_{Sun}$)\\
$\gamma $ & 37.20 & 37.18 & $\pm$ 0.23 & Systemic radial velocity (km s$^{-1}$)\\
$\gamma_{WD HRS}$ & 44.5 & 44.4 & $\pm$ 3.4 & RV offset in HRS (km s$^{-1}$)\\
$\gamma_{WD STIS}$ & -1.1 & 0.1 & $\pm$ 3.0 & RV offset in STIS (km s$^{-1}$)\\
\hline
\multicolumn5c{Derived parameters} \\
\hline
$a$ & 0.01496 & 0.01501 & $\pm$ 0.00013 & Orbital semi-major axis (au) \\
$i$ & 80.8 & 80.9 & $\pm$ 1.2 & Orbital inclination (deg) \\
$M_{WD}$ & 0.792 & 0.797 & $\pm$ 0.016 & White-dwarf mass ($M_{Sun}$) \\
$R_{WD}$ & 0.01134 & 0.01140 & $\pm$ 0.00059 & White-dwarf radius in ($R_{Sun})$ \\
$R_{WD, Ein}$ & 0.004647 & 0.004671 & $\pm$ 0.000066 & White-dwarf Einstein radius ($R_{Sun}$) \\
log(g) & 8.227 & 8.225 & $\pm$ 0.048 & White-dwarf log(g) (cm s$^2$) \\
$M_{MS}$ & 0.853 & 0.864 & $\pm$ 0.029 & K dwarf mass ($M_{Sun}$) \\
$R_{MS}$ & 0.816 & 0.816 & $\pm$ 0.042 & K dwarf radius ($R_{Sun}$) \\
\enddata
\end{deluxetable*}

\section{Discussion} \label{sec:discussion}

\begin{figure}
    \centering
    \includegraphics[width=0.45\textwidth]{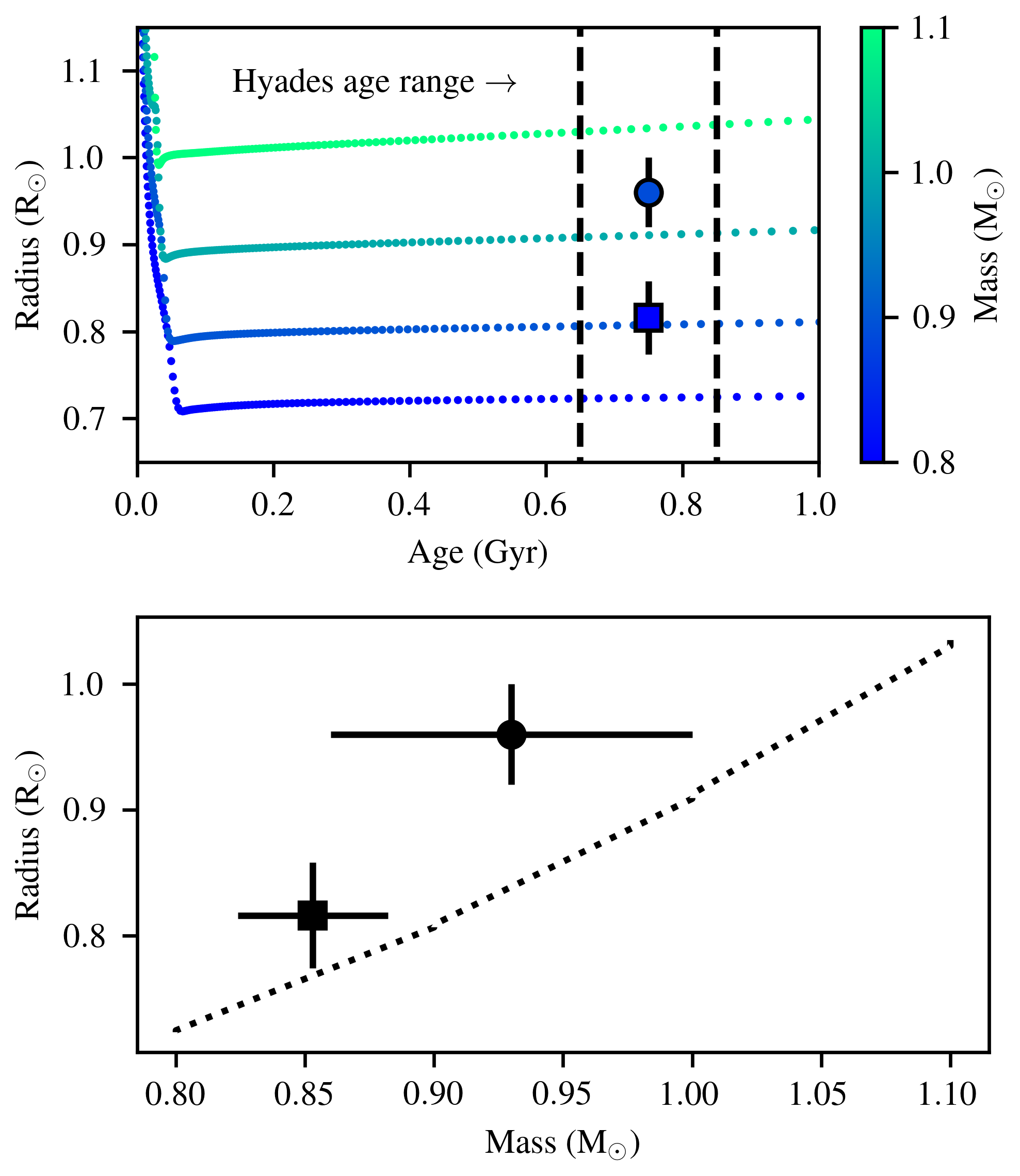}
    \caption{Measurements for the radius and mass of the main-sequence component of V471 Tau (circle, square) compared to MIST evolutionary models (assuming [Fe/H]=0.0).  Top: Radius vs. age, colored by mass.  Bottom: Radius vs. mass, showing the MIST model for ages between 650 and 850 Myr.  The dots indicate the MIST model values, the square indicates our measurement for the main-sequence star, and the circle indicates a previous measurement from the literature \citep{OBrien2001}.}
    \label{fig:ms_evol}
\end{figure}

Figure \ref{fig:ms_evol} shows the age and radius of the K dwarf compared to evolutionary models across the age of the Hyades.  \citet{OBrien2001} found that the dK star radius is larger than predictions from stellar-evolutionary models.  They propose that either the star is out of thermal equilibrium due to the recent immersion in a common envelope, or that the star is larger due to reduced convective efficiency from strong magnetic fields, similar to what has been seen in some, but not all, dM eclipsing binary stars \citep[e.g.][]{Mullan2001, Chabrier2007, Han2019} and rapidly rotating single dM stars \citep[][]{Kesseli2018}.  However, in our analysis, we find that the dK star is consistent with single-star stellar-evolutionary models.

An estimate for the maximum mass that could be accreted during the CE phase is given by $M_{\rm acc}\sim \dot{M}_{\rm Edd}\times t_{CE}$ where $t_{CE}$ is the orbital decay timescale and $\dot{M}_{\rm Edd}$ is the Eddington-limited accretion rate.\footnote{$\dot{M}_{\rm Edd}\sim 10^{-3}\left(R_\star/R_\odot\right) M_\odot yr^{-1}$ for main-sequence stars.} Since the orbital decay timescale during the CE is short ($<$ 1 yr; \citealt{2019MNRAS.485.4492W}), the maximum mass that the dK star could have accreted is only 0.1\%, thereby providing further support that its evolution is consistent with single-star evolution models.

\begin{figure}
    \centering
    \includegraphics[width=0.47\textwidth]{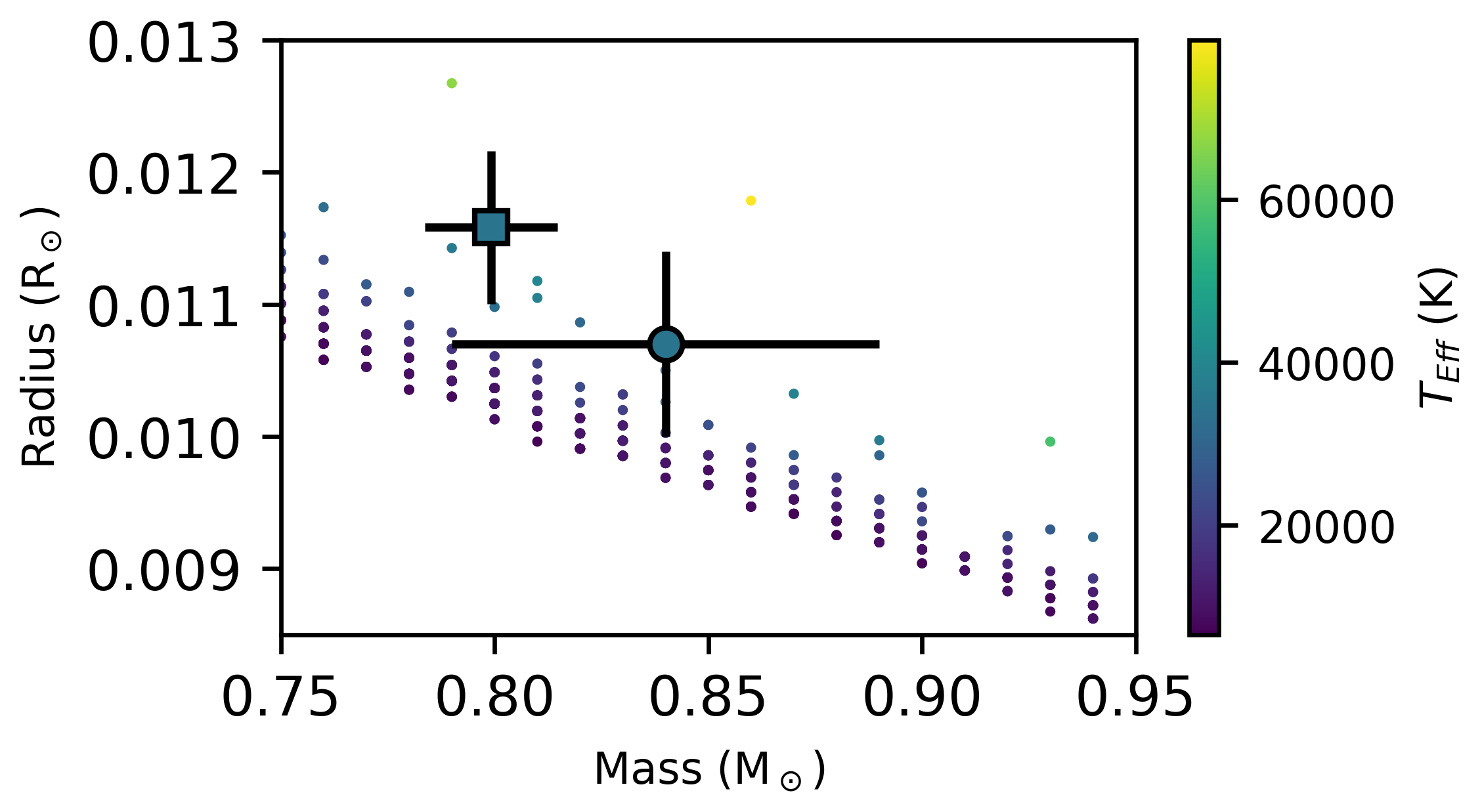}
    \caption{Radius vs. mass for white dwarfs and the white dwarf component of V471 Tau.  Small circles show measurements from the Sloan Digital Sky Survey \citep{Tremblay2011}.  The square indicates our measurement for the mass and radius of the main-sequence star, and the circle indicates a previous measurement from the literature \citep{OBrien2001}.}
    \label{fig:wd_mass_rad}
\end{figure}

Figure \ref{fig:wd_mass_rad} shows the white-dwarf mass and radius compared to \citet{OBrien2001} and white dwarfs from the Sloan Digital Sky Survey \citet[][]{Tremblay2011}. We note that our derived value for the surface gravity of the white dwarf, log(g)=8.227$\pm$0.048, is consistent with values determined by \citet{Werner1997} and \citet{Barstow1997} using the broadening of the Lyman series absorption lines (8.21 $\pm$ 0.23 and 8.16 $\pm$ 0.18, respectively), a completely independent method.  As discussed in Section \ref{sec:intro}, the high temperature and mass of the white dwarf is inconsistent with single-star evolution given the age of the Hyades.  A plausible formation scenario for V471 Tau is that initially, it was a hierarchical triple in which the inner binary merged when both stars were on the main-sequence, as proposed by \citet{OBrien2001}.  As the merged product evolved off the main-sequence, it entered a common envelope with the K-type main-sequence star.  The orbit subsequently decayed during the CE until the envelope was ejected leaving the system in its current short-period configuration \citep{2019MNRAS.485.4492W}.

To determine the initial mass range of the merged progenitor, we use the initial-final mass relationship (IFMR) derived from white dwarfs in clusters \citep{2018ApJ...866...21C}\footnote{A caveat is that even if all the mass remains bound during and after the merger of the inner binary, the resulting star may not follow the observationally derived IMFR relationship.}.  Including both the uncertainty in the mass of the white dwarf and the uncertainty in the initial-final-mass relation, we obtain a progenitor mass range of 3.16-3.41 $M_\odot$.  We model the evolution of stars in this mass range using the ``Modules for Experiments in Stellar Evolution" code (version 8845; \citealt{2011ApJS..192....3P,2013ApJS..208....4P}).  Because V471 Tau is a member of the Hyades, we employed a metallicity of $[{\rm Fe/H}]=+0.15$, consistent with the cluster.  Mass loss on the Red Giant Branch (RGB) followed a Reimer’s prescription while mass-loss on the Asymptotic Giant Branch (AGB) followed a Bloecker prescription \citep{1975MSRSL...8..369R,1995A&A...297..727B}.  Both mass-loss prescriptions require a mass-loss coefficient, which we took as $\eta_{\rm R}=0.4$ on the RGB and $\eta_{\rm B}=0.4$ on the AGB.  We choose these values such that the final white-dwarf mass that emerges is consistent with the observationally derived initial-final mass relationship \citep{2018ApJ...866...21C}.

Since the K-type star entered a common envelope, we can determine the minimum initial orbital separation required to avoid such a fate in a two-body system assuming an initially circular orbit\footnote{Note that this scenario assumes that no mass was lost during the merger of the inner binary. Any mass lost from the system during the merger of the two main-sequence stars, would widen the resultant binary, thereby requiring an even closer initial separation to enter a common envelope.} \citep{2010MNRAS.408..631N}.  Because the initial mass ratio of the binary, $q\equiv M_{\rm WD}/M_{\star}$, ranges from 0.23 to 0.25, tidal torques will synchronize the system implying that the minimum semi-major axis required to escape the CE phase is just outside the maximum radius during the primary's evolution \citep{Nordhaus2013}.  A reasonable estimate for the maximum value would be 1.9 au, i.e. ten percent larger then the maximum radius obtained from the evolution of all models.  If the K dwarf were orbiting just exterior to that separation, it would avoid common envelope such that the orbital separation widens. In this situation, the orbit expands and reaches a final semi-major axis that can be estimated as $a_{\rm}\sim R_{\rm MS, max}\times (M_{\star}/M_{\rm WD})$ where $M_{\star}$ is the initial mass of the primary star (i.e. the zero-age-main-sequence mass for the white-dwarf progenitor; \citealt{Nordhaus2013}).  If the K dwarf had avoided a CE phase, it would have to have had an orbital separation greater than $\sim 2.7$ au. In summary, the initial orbital separation of the system had to be less than 1.9 au to enter a CE and emerge in its current position, otherwise the orbit would expand to at least 2.7 au.

\section{Conclusions}\label{sec:conclusions}

In this paper, we presented results of an investigation into the V471 Tau post-common envelope system, using archival data from NASA's K2 Mission and the HST.  We fitted an eclipse model employing a Gaussian processes approach to capture the modulation of the light curve due to rotation of the main-sequence star.  In our fit, we recovered the transit of the white dwarf across the main sequence star for the first time, including the expected filling in of the transit due to gravitational microlensing.  We independently analyzed HST-STIS observations of the V471 Tau to obtain additional radial-velocity measurements of the white dwarf star.  Combining all of the measurements, we measured a mass and radius for the white dwarf that is generally consistent with previous investigations, but with higher precision.  Unlike previous investigations, we measured a mass and radius for the main-sequence star that is consistent with single-star evolution given the age and metallicity of the Hyades.

We calculate a progenitor mass of 3.16-3.41 $M_\odot$ for the white dwarf.  Like previous investigations, we find that this is too high a mass given the white dwarf's temperature and age of the Hyades, indicating that it might be the product of a merger scenario.  We find that the initial orbital separation of the main-sequence star and merged white-dwarf progenitor must have been less than 1.9 au in order to result in the system as we see it today.

\begin{acknowledgments}
We thank the anonymous referee for the helpful comments that improved the manuscript.  We thank JJ Hermes and Todd Vaccaro for their thoughtful correspondence.  This material is based upon work supported by the National Science Foundation under grant No. 2009713.  This paper includes data collected by the Kepler mission and obtained from the MAST data archive at the Space Telescope Science Institute (STScI). Funding for the Kepler mission is provided by the NASA Science Mission Directorate. STScI is operated by the Association of Universities for Research in Astronomy, Inc., under NASA contract NAS 5–26555.  This research includes observations made with the NASA/ESA Hubble Space Telescope obtained from the Space Telescope Science Institute, which is operated by the Association of Universities for Research in Astronomy, Inc., under NASA contract NAS 5–26555.  
\end{acknowledgments}

\vspace{5mm}
\facilities{HST(STIS), Kepler}

\software{astropy \citep{2013A&A...558A..33A,2018AJ....156..123A},  \texttt{celerite} \citep{Foreman-Mackey2017}, \texttt{corner} \citep{corner}, \texttt{eb} \citep{Irwin2011}, \texttt{emcee} \citep{Foreman-Mackey2013}, \texttt{everest} \citep{Luger2016, Luger2018}}, \texttt{MESA} \citep{2011ApJS..192....3P,2013ApJS..208....4P}

\bibliography{references}{}
\bibliographystyle{aasjournal}

\end{document}